\documentclass[final,5p,times,twocolumn]{elsarticle}

\usepackage{graphicx}
\usepackage[pdftex,bookmarks]{hyperref}
\usepackage{amssymb}
\usepackage{gensymb}
\usepackage{units}
\usepackage{booktabs}
\usepackage{multirow}
\usepackage{lineno}
\journal{Nuclear Instruments and Methods A}

\begin{document}

\begin{frontmatter}

%% Title, authors and addresses

%% use the tnoteref command within \title for footnotes;
%% use the tnotetext command for theassociated footnote;
%% use the fnref command within \author or \address for footnotes;
%% use the fntext command for theassociated footnote;
%% use the corref command within \author for corresponding author footnotes;
%% use the cortext command for theassociated footnote;
%% use the ead command for the email address,
%% and the form \ead[url] for the home page:
%% \title{Title\tnoteref{label1}}
%% \tnotetext[label1]{}
%% \author{Name\corref{cor1}\fnref{label2}}
%% \ead{email address}
%% \ead[url]{home page}
%% \fntext[label2]{}
%% \cortext[cor1]{}
%% \address{Address\fnref{label3}}
%% \fntext[label3]{}

\title{The upgrade of the ALICE Inner Tracking System}

% if there is only one institution, use this form:
\author[iphc]{Serhiy Senyukov for the ALICE-ITS collaboration}
\ead{serhiy.senyukov@cern.ch}
\address[iphc]{Universit\'{e} de Strasbourg, IPHC-CNRS, 23 rue du Loess 67037 Strasbourg, France}

\begin{abstract}
ALICE is a general purpose experiment dedicated to the study of nucleus-nucleus collisions at LHC. After more than 3 years of successful operation, an upgrade of the apparatus during the second long shutdown of LHC (LS2) in 2017/18 is in preparation. One of the major goals of the proposed upgrade is to extend the physics reach for rare probes at low transverse momentum. The reconstruction of the rare probes requires a precise determination of the primary and secondary vertices that is performed in ALICE by the Inner Tracking System (ITS). The present ITS made of 6 layers of three technologies of silicon devices allows, for example, to reconstruct $D$ mesons with the transverse momentum down to $\sim \unit[2]{GeV/c}$.

Further extension of this range towards lower $p_T$ requires the installation of the new ITS consisting of 7 layers of silicon detectors with significantly better single point resolution and reduced material budget. It is expected that the new system will allow to improve the impact parameter resolution by a factor of $\sim 3$. Moreover, the data rate capability of the upgraded ITS should be significantly improved in order to exploit the full expected LHC lead-lead interaction rate of \unit[50]{kHz}, almost two orders of magnitude above the present readout rate.

The present contribution describes first the requirements for the new ITS followed by the conceptual design of the system and its expected performance. Secondly, an overview of the different R\&D activities from the concept towards the final detector is given.
\end{abstract}

\begin{keyword}
Heavy ions, ALICE, ITS, upgrade, CMOS pixel sensors

%% PACS codes here, in the form: \PACS code \sep code
%% Find PACS codes here: http://www.aip.org/pacs/pacs2010/individuals/pacs2010_regular_edition/index.html

%% MSC codes here, in the form: \MSC code \sep code
%% or \MSC[2008] code \sep code (2000 is the default)

\end{keyword}

\end{frontmatter}

%% \linenumbers

%% main text
\section{Introduction}
ALICE (A Large Ion Collider Experiment)\cite{ALJINST} is one of the four main LHC experiments. Its main purpose is to study the properties of the Quark-Gluon Plasma (QGP) created in the ultra-relativistic collisions of heavy ions. This study is performed via various probes emerging from the collisions. Rare probes like charmed and beauty mesons and baryons represent a powerful tool giving valuable information about the strong interactions inside the QGP. Suppression\cite{RAA_D} and elliptic flow\cite{Flow_D} of the $D$ mesons in Pb-Pb collisions has been measured by ALICE down to $p_T =\unit[2]{GeV/c}$. Reaching lower transverse momentum values seems to be precluded with the current setup, due to the large combinatorial background level. 

Reconstruction of the charmed mesons down to zero transverse momentum will allow to asses the degree of thermalization of charm quarks and even possible thermal production of charm. Thus, extending the limits of the open charm reconstruction down to zero $p_T$ became one of the major motivations for the future ALICE upgrade. Other probes that would become available with the upgraded detector are beauty mesons and baryons, quarkonia, low-mass dileptons and heavy nuclear states.

\section{General ALICE upgrade}
The general ALICE upgrade strategy aiming to improve the physics reach of the experiment has been developed in 2011--2012. The performance gain will be achieved thanks to two important improvements. The first one is the increased luminosity of the LHC in Pb-Pb collisions after the second Long Shutdown (LS2) reaching $\mathcal{L}=6 \times 10^{27} cm^{-2}s^{-1}$ and resulting in the minimum bias event rate of about \unit[50]{kHz}\cite{LHC_HI_LS2}. The second one is an upgrade  of the main tracking detectors at central rapidity.
The general ALICE upgrade strategy includes several projects:
\begin{enumerate}
\item Reduction of the beam pipe radius from the present value of $r_{out}=\unit[29.8]{mm}$ down to $\unit[20]{mm}$. This modification will allow to place the first tracking layer much closer to the interaction point ($r_1\simeq\unit[22]{mm}$ instead of \unit[34]{mm} in the present setup);
\item Installation of the new Inner Tracking System (ITS) featuring higher granularity and lower material budget; 
\item Upgrade of the Time Projection Chamber (TPC)\cite{TPC_VCI2013} consisting in the replacement of the wire chambers with GEM detectors and new pipelined electronics allowing for the high readout rate; 
\item Upgrade of the readout electronics of the Transition Radiation (TRD) and Time-Of-Flight (TOF) detectors, Photon (PHOS) and Muon spectrometers for the high readout rate;
\item Upgrade of the forward trigger detectors: V0, T0 and the Zero Degree Calorimeters (ZDCs);
\item Upgrade of the online systems and offline reconstruction and analysis framework. 
\end{enumerate}
More details on the general ALICE upgrade strategy can be found in the Letter Of Intent\cite{ALICE_upgrade_LOI} endorsed by LHCC in September 2012.

\section{ALICE ITS upgrade requirements}
The construction of the new ITS is an important part of the general ALICE upgrade. Indeed, the main purpose of the ITS is to improve the primary vertex position and reconstruct the decay vertices of heavy flavour particles. An improvement of the spatial resolution of the ITS by increasing the segmentation and decreasing the material budget will allow to extend the accessible transverse momentum range well below $p_T=\unit[2]{GeV/c}$. The new ITS will have to cope with the important event rates of \unit[50]{kHz} in Pb-Pb collisions and several hundreds of kHz in pp collisions. Such event rates will lead to a relatively high radiation load on the detector. According to recent simulations, the overall dose expected for the full physics program after LS2 for the innermost layer can reach up to \unit[700]{kRad} and \unit[$10^{13}$]{$n_{eq}/cm^2$} including a safety factor of 4.
\section{Conceptual design and expected performance of the upgraded ITS}
The work on the conceptual design of the new ITS was started in 2011 by the investigation of possible technologies for the new detector. The following three technological options were initially considered.
\begin{description}
\item[Hybrid pixel detectors] represent a well-known technology with proven radiation hardness. Hybrid pixel detectors were used to equip the two innermost layers of the present ALICE ITS, as well as the tracking detectors of CMS and ATLAS experiments. However, this technology has some important limitations. First, the pixel pitch is limited to about $\unit[50]{\mu m}$ by the bump bonding process and second the cost of the flip-chip bonding does not allow to equip a large surface with this type of detectors. 

\item[CMOS pixel detectors] is a novel approach to the silicon particle detectors. CMOS process allows to integrate the sensing volume and the readout electronics in a single silicon die. Such an integration gives a possibility to reduce significantly the pixel pitch to about $\unit[20]{\mu m}$. Moreover, by the thinning of the chip to $\sim \unit[50]{\mu m}$ one can substantially reduce the silicon contribution to the overall material budget. The radiation hardness of CMOS sensors was limiting their application in high energy physics experiments. Nevertheless, recent developments in CMOS technology allowed to significantly improve the tolerance both to the ionizing and non-ionizing radiations. Presently the STAR\cite{STAR_overview} collaboration is installing \emph{PXL}\cite{STAR-PXL} -- the first tracking system based on CMOS sensors. It is equipped with 400 \emph{Ultimate}\cite{MIMOSA28-STAR} sensors produced in the \emph{AMS 0.35 $\mu m$} process that can withstand the combined radiation dose of \unit[150]{kRad} and \unit[$3 \times 10^{12}$]{$n_{eq}/cm^2$} at an operation temperature of \unit[30] {\celsius}.

\item[Micro-strip detectors] represent another well-known and widely used technology. An important added value of this technology is the possibility of the particle identification (PID) via the specific ionizing energy-loss $dE/dx$ in the silicon. The micro-strip detector for the upgrade can have 1536 strips on each side of the \unit[300]{$\mu m$} sensor. Strip pitch can be \unit[95]{$\mu m$} with the stereo angle between the strips on opposite sides of \unit[35]{mrad}. The limitations of micro-strip detectors are the limited spatial resolution along the beam axis and the low granularity. The low granularity makes them usable only in the low track density conditions i.e. in the outermost layers of the ITS.
\end{description}

\begin{figure}
\centering
\includegraphics[width=1.\linewidth]{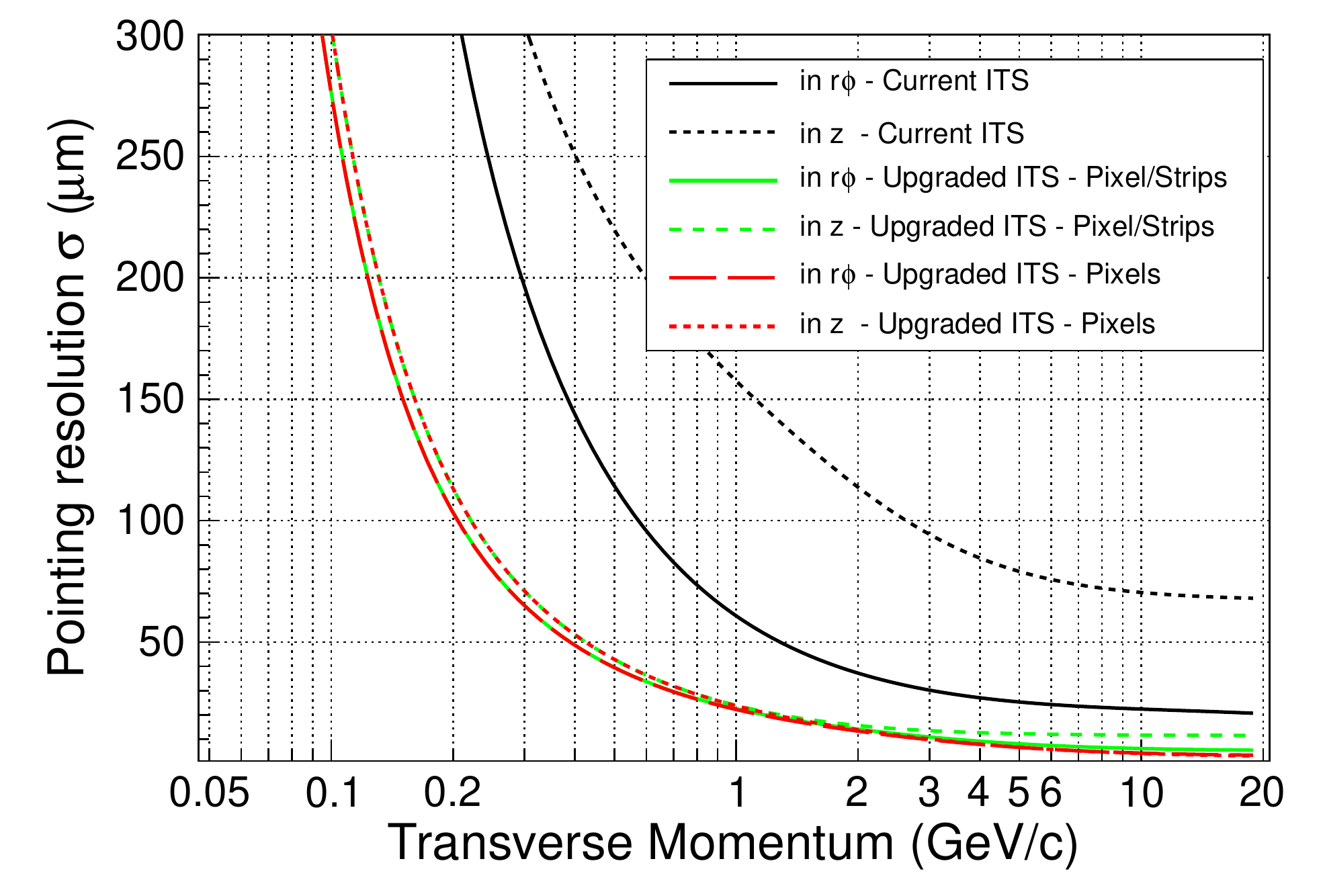}
\caption{Pointing resolution to the vertex for charged pions as a function of $p_T$ for the current ITS and the upgraded ITS. Ref. \cite{ALICE_ITS_CDR}}
\label{fig:PointingResolution}
\end{figure}

After considering pros and cons of the available technological solutions a conceptual design of the new ITS was proposed. It includes 7 layers of pixel detectors. Optionally the 4 outermost layers can be equipped with the silicon micro-strip detectors. The parameters of the proposed layout are summarized in \autoref{tab:concept_layouts}. The intrinsic resolutions and material budget of the pixel layers correspond to what can be reached with the CMOS pixel sensors. The longitudinal extensions of the new layers were chosen to provide the pseudo-rapidity coverage of $|\eta|<1.22$ over \unit[90]{\%} of the luminous region. The radial positions of the layers were tuned to obtain the optimal combined performance in terms of the pointing resolution, $p_T$-resolution and tracking efficiency. The expected pointing resolution of the new layout is compared with that of the present ITS in \autoref{fig:PointingResolution}. It is clear that the ITS upgrade can significantly improve the performance of the detector. For example, the pointing resolution in $r\phi$ plane becomes 3 times better at \unit[1]{$GeV/c$}, passing from about $\unit[60]{\mu m}$ to $\simeq \unit[20]{\mu m}$.
%\begin{figure}
%\centering
%\includegraphics[width=1.\linewidth]{./PtResolution}
%\caption{Realtive transverse momentum resolution for charged pions as a function of $p_T$ for the current ITS and the two conceptual layouts.}
%\label{fig:PtResolution}
%\end{figure}
%\begin{figure}
%\centering
%\includegraphics[width=1.\linewidth]{./TrackingEfficiency}
%\caption{Stand-alone tracking efficiency for charged pions as a function of $p_T$ for the current ITS and the two conceptual layouts.}
%\label{fig:TrackingEfficiency}
%\end{figure}

\begin{table*}
\caption{Parameters of the conceptual layout of the ITS upgrade. Numbers in parentheses correspond to the micro-strip detectors.}
\label{tab:concept_layouts}
\centering
\begin{tabular}{ccccccc}
\toprule
\multirow{2}{*}{Layer} & \multirow{2}{*}{Type} & \multirow{2}{*}{Radius [cm]} & \multirow{2}{*}{Length [cm]}& \multicolumn{2}{c}{Intrinsic resolution $[\mu m]$} & \multirow{2}{*}{Material budget [\% $X_0$]}\\
&&&&$r\phi$&$Z$&\\
\midrule
Beam pipe & - & 2.0 & - & - & - &0.22\\
\midrule
1 & \multirow{3}{*}{Pixels} & 2.2 & 22.4 & 4 & 4 & 0.3\\
2 & & 2.8 & 24.2 & 4 & 4 & 0.3\\
3 & & 3.6 & 26.8 & 4 & 4 & 0.3\\
\midrule
4 & \multirow{4}{*}{Pixels (Strips)} & 20.0 & 78.0 & 4 (20) & 4 (830) & 0.3 (0.83)\\
5 & & 22.0 & 83.6 & 4 (20) & 4 (830) & 0.3 (0.83)\\
6 & & 41.0 & 142.4 & 4 (20) & 4 (830) & 0.3 (0.83)\\
7 & & 43.0 & 148.6 & 4 (20) & 4 (830) & 0.3 (0.83)\\
\bottomrule
\end{tabular}
\end{table*}

The conceptual design of the new ITS was finalized by the end of August 2012. It was described in the corresponding report\cite{ALICE_ITS_CDR} that was reviewed by the LHCC in September 2012. Following this review, LHCC endorsed the ITS upgrade project and encouraged the ALICE collaboration to prepare the Technical Design Report.

\section{R\&D towards the final detector design}
A dedicated and intense research and development activity is being carried out now in order to evaluate the technical feasibility of the upgrade and to develop the final detector design. It covers various topics like the pixel chip development, mechanical layout design, cooling system and final detector integration.

At present the selected technology for the pixel sensor is \emph{TowerJazz}\footnote{\url{http://www.jazzsemi.com/}} CIS process.

\subsection{Pixel chip development}
Following the laboratory and beam tests of the MIMOSA 32 prototype chips in 2012, the radiation hardness of the \emph{TowerJazz \unit[0.18]{$\mu m$}} imaging CMOS process has been validated up to the combined load of \unit[1]{MRad} and \unit[$10^{13}$] {$n_{eq}/cm^2$} at \unit[30] {\celsius} \cite{Senyukov_RESMDD2012}.  

Three groups are presently working in parallel on the design of the final pixel chip. The main challenge of this development is to satisfy simultaneously the stringent requirements in terms of the spatial resolution, readout speed and power consumption. The PICSEL group of IPHC (Strasbourg) is working on the baseline chip called MISTRAL. It will be based on the rolling shutter readout with column level discriminators providing the full integration time of \unit[30]{$\mu s$}. The expected power density is about \unit[400]{$mW/cm^2$}. Additionally, the PICSEL group works on a more advanced chip called ASTRAL. The latter will have in-pixel discriminators that will allow to reduce the integration time to \unit[15]{$\mu s$} and the power density to \unit[350]{$mW/cm^2$}. More details on these developments can be found in Ref.\cite{Baudot_VCI2013}. Two other groups from RAL (UK) and CERN are working on the further optimization of the architecture in order to reduce the integration time and power consumption of the chip. In the approach being developed in RAL the pixel matrix will be divided into several sub-arrays that will be read out in the parallel rolling shutter mode. An architecture proposed by the CERN group relies on the in-pixel discriminators and the readout using the priority encoder technique. 
\subsection{Mechanical design and cooling}
The mechanical layout of the new ITS has been divided in 2 independent parts: Inner Barrel and Outer Barrel. Due to the different mechanical constraints the design of the two barrels is being done separately.
\subsubsection{Inner barrel design}
The Inner Barrel will include the 3 innermost layers. Each layer is composed of the staves containing 9 pixel chips of the size of $\unit[15\times 30]{mm^2}$ and \unit[50]{$\mu m$} thickness. The relevant dimensions and composition of the inner barrel are summarized in \autoref{tab:InnerBarrel_layout}.
%The front section of the barrel is shown on \autoref{fig:InnerBarrelFront}. 

\begin{figure}
\centering
\includegraphics[width=1\linewidth]{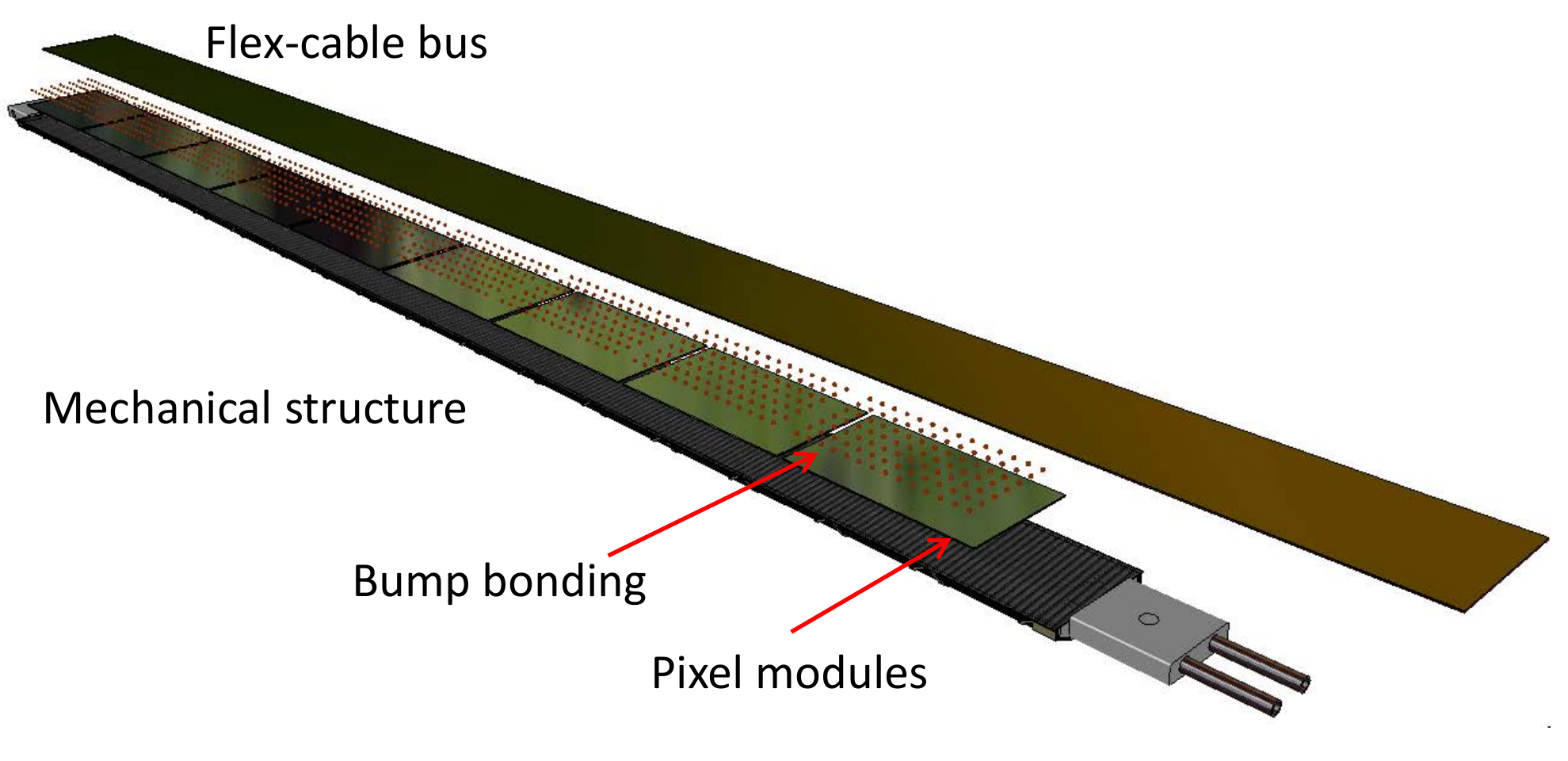}
\caption{Composition of the stave. Ref. \cite{ALICE_ITS_CDR}}
\label{fig:stave}
\end{figure}

\autoref{fig:stave} shows the preliminary composition of the stave. Nine chips will be first flip-chip bonded to the aluminium-polyimide flex-cable with BGA balls or via laser soldering. Then the cable with the chips will be glued to the carbon fibre support which embeds the water cooling circuit. Several solutions are under consideration for the cooling including polyimide tubes, polyimide micro-channels planes and silicon micro-channels planes. More details on the mechanical structure design and the cooling options can be found in Ref.\cite{Feofilov_VCI2013}.

\begin{figure}
\centering
\includegraphics[width=1\linewidth]{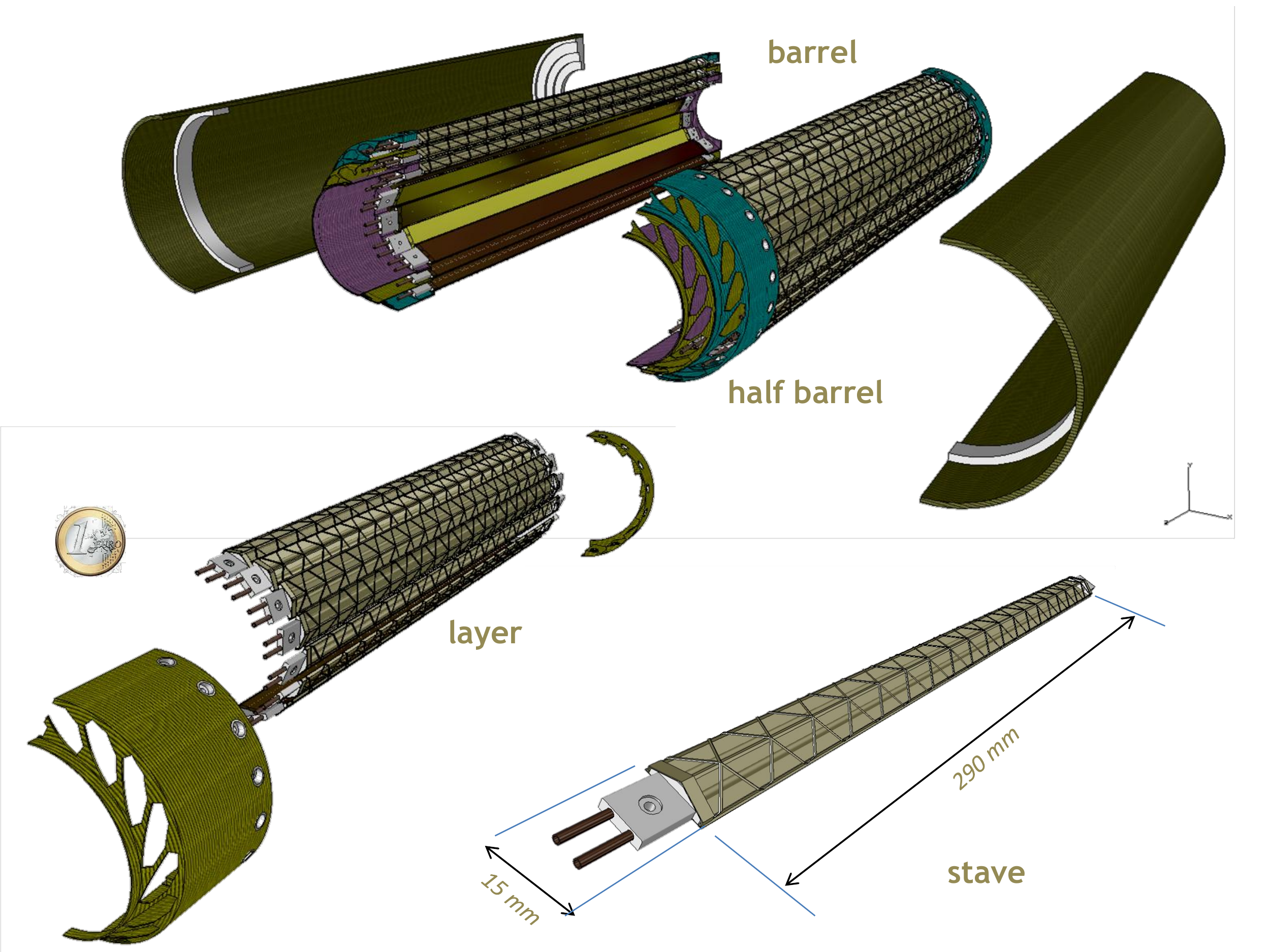}
\caption{Assembly of the Inner Barrel. Ref. \cite{ALICE_ITS_CDR}. The inner barrel composed of the two half-barrels and a cylindrical outer support shell is shown at the top. A part of the half-barrel corresponding to one of the layers composed of the staves and supporting carbon structures is shown at the bottom.}
\label{fig:InnerBarrelAssembly}
\end{figure}

The assembly scheme for the Inner Barrel is schematically shown on \autoref{fig:InnerBarrelAssembly}. The top part of the figure shows the Inner Barrel composed of the two half-barrels and a cylindrical outer support shell. The bottom part of the figure shows a part of the half-barrel corresponding to one of the layers. This part is formed by the staves mounted on the two carbon half-wheels.

First prototypes of the Inner Barrel and staves have been produced at CERN in 2012 proving the technical feasibility of this design.
\begin{table*}
\caption{Mechanical design of the inner barrel.}
\label{tab:InnerBarrel_layout}
\centering
\begin{tabular}{cccccc}
\toprule
Layer & Radius [mm] & Stave length [mm]& Staves & Chips/stave & Material budget [\% $X_0$]\\
\midrule
1 & 22 & \multirow{3}{*}{270} & 12 & \multirow{3}{*}{9} & 0.3\\
2 & 28 &  & 16 &  & 0.3\\
3 & 36 &  & 20 &  & 0.3\\
%\midrule
%4 & 200 & \multirow{2}{*}{843} & 48 & \multirow{2}{*}{28} & 0.8\\
%5 & 220 &  & 52 &  & 0.8\\
%\midrule
%6 & 410 & \multirow{2}{*}{1475} & 96 & \multirow{2}{*}{52} & 0.8\\
%7 & 430 & & 102 &  & 0.8\\
\bottomrule
\end{tabular}
\end{table*}
%\begin{figure}
%\centering
%\includegraphics[width=1\linewidth]{./InnerBarrelFront}
%\caption{Transversal view of the inner barrel.}
%\label{fig:InnerBarrelFront}
%\end{figure}

\subsubsection{Outer barrel design}
The design studies for the outer barrel are still in progress. This is mostly due to the pending choice between pixel and micro-strip detectors for the 4 outermost layers. This choice will define the mechanical constraints for the final layout. Presently the pixel option is being considered as the baseline and the final decision will be taken in 2013. Several aspects will be taken into account:
\begin{itemize}
\item Possible benefits of the PID provided by the micro-strip detectors for some specific physics analyses;
\item Technical feasibility of the large area ($\sim \unit[10]{m^2}$) pixel layers in terms of the power distribution, cooling and mechanical integration;
\item Overall cost of the two options.
\end{itemize}

\subsection{Final integration}
\begin{figure}
\centering
\includegraphics[width=1\linewidth]{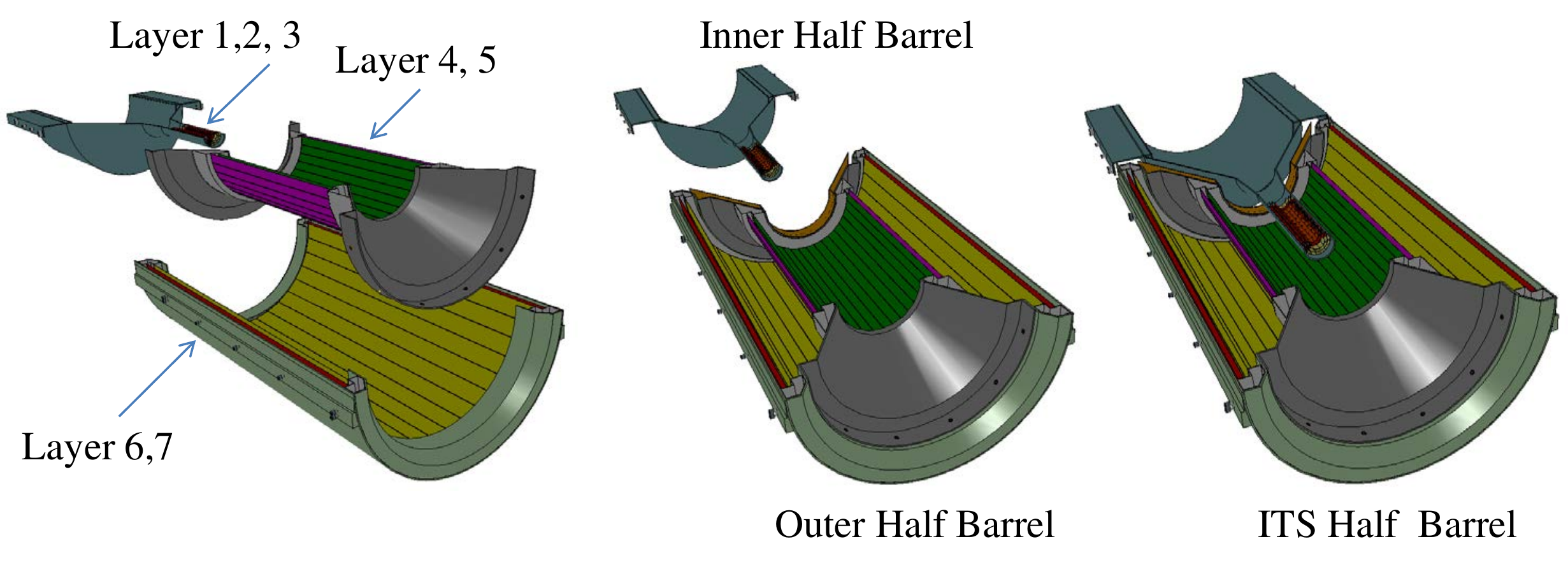}
\caption{Final integration of the detector. Ref. \cite{ALICE_ITS_CDR}}
\label{fig:Integration}
\end{figure}

The final integration of the ITS is schematically shown on \autoref{fig:Integration}. The integration procedure is designed in the way to permit the installation and removal of the ITS without moving the TPC. This opens the possibility of the detector maintenance during the short winter shutdown of the LHC. 

\section{Conclusions}
A new upgraded ITS is expected to significantly improve the physics performance of the ALICE detector in reconstruction of the rare probes at low transverse momentum. It will be reached by relying on CMOS pixel sensors that guarantee the high granularity and low material budget. The radiation hardness of the \emph{TowerJazz \unit[0.18]{$\mu m$}} CMOS process has been validated up to the level of \unit[1]{MRad} and \unit[$10^{13}$] {$n_{eq}/cm^2$} at \unit[30] {\celsius}. Presently an intense R\&D for the final pixel chip is in progress. 

First mechanical and cooling prototypes showed encouraging results in terms of the material budget and heat removing properties, proving the feasibility of the low mass ITS with the material budget of $\sim \unit[0.3]{\% X_0}$ per layer.

\bibliographystyle{elsarticle-num}
\bibliography{Senyukov-VCI2013}

\begin{thebibliography}{10}
\expandafter\ifx\csname url\endcsname\relax
  \def\url#1{\texttt{#1}}\fi
\expandafter\ifx\csname urlprefix\endcsname\relax\def\urlprefix{URL }\fi
\expandafter\ifx\csname href\endcsname\relax
  \def\href#1#2{#2} \def\path#1{#1}\fi

\bibitem{ALJINST}
K.~Aamodt, et~al., {The ALICE experiment at the CERN LHC}, JINST 0803 (2008)
  S08002.
\newblock \href {http://dx.doi.org/10.1088/1748-0221/3/08/S08002}
  {\path{doi:10.1088/1748-0221/3/08/S08002}}.

\bibitem{RAA_D}
B.~Abelev, et~al., {Suppression of high transverse momentum D mesons in central
  Pb-Pb collisions at $\sqrt{s_{NN}}=2.76$ TeV}, JHEP 1209 (2012) 112.
\newblock \href {http://arxiv.org/abs/1203.2160} {\path{arXiv:1203.2160}},
  \href {http://dx.doi.org/10.1007/JHEP09(2012)112}
  {\path{doi:10.1007/JHEP09(2012)112}}.

\bibitem{Flow_D}
D.~Caffarri, {Measurement of the $D$ meson elliptic flow in Pb-Pb collisions at
  $\sqrt{s_{NN}}=2.76$ TeV with ALICE }\href {http://arxiv.org/abs/1212.0786}
  {\path{arXiv:1212.0786}}.

\bibitem{LHC_HI_LS2}
D.~Manglunki, \href{http://cds.cern.ch/record/1493028}{Plans for ions in the
  injector complex} (2012) 6 p.
\newline\urlprefix\url{http://cds.cern.ch/record/1493028}

\bibitem{TPC_VCI2013}
B.~Ketzer, {A Time Projection Chamber for High-Rate Experiments: Towards an
  Upgrade of the ALICE TPC}, 2013.
\newblock \href {http://arxiv.org/abs/1303.6694} {\path{arXiv:1303.6694}}.

\bibitem{ALICE_upgrade_LOI}
{ALICE Collaboration}, \href{http://cds.cern.ch/record/1475243}{{Letter of
  Intent for the Upgrade of the ALICE Experiment}}, Tech. Rep.
  CERN-LHCC-2012-012. LHCC-I-022, CERN, Geneva (Aug 2012).
\newline\urlprefix\url{http://cds.cern.ch/record/1475243}

\bibitem{STAR_overview}
K.~Ackermann, et~al., {STAR detector overview}, Nucl.Instrum.Meth. A499 (2003)
  624--632.
\newblock \href {http://dx.doi.org/10.1016/S0168-9002(02)01960-5}
  {\path{doi:10.1016/S0168-9002(02)01960-5}}.

\bibitem{STAR-PXL}
L.~Greiner, E.~Anderssen, H.~Matis, H.~Ritter, J.~Schambach, et~al., {A MAPS
  based vertex detector for the STAR experiment at RHIC}, Nucl.Instrum.Meth.
  A650 (2011) 68--72.
\newblock \href {http://dx.doi.org/10.1016/j.nima.2010.12.006}
  {\path{doi:10.1016/j.nima.2010.12.006}}.

\bibitem{MIMOSA28-STAR}
A.~Besson, M.~Winter, J.~Baudot, G.~Bertolone, N.~Chon-Sen, et~al., {CMOS
  sensors with high resistivity epitaxial layer}, PoS EPS-HEP2011 (2011) 204.

\bibitem{ALICE_ITS_CDR}
{ALICE Collaboration}, \href{http://cds.cern.ch/record/1475244}{{Conceptual
  Design Report for the Upgrade of the ALICE ITS}}, Tech. Rep.
  CERN-LHCC-2012-013. LHCC-P-005, CERN, Geneva (2012).
\newline\urlprefix\url{http://cds.cern.ch/record/1475244}

\bibitem{Senyukov_RESMDD2012}
S.~Senyukov, J.~Baudot, A.~Besson, G.~Claus, L.~Cousin, et~al., {Charged
  particle detection performances of CMOS pixel sensors produced in a 0.18 um
  process with a high resistivity epitaxial layer }\href
  {http://arxiv.org/abs/1301.0515} {\path{arXiv:1301.0515}}, \href
  {http://dx.doi.org/10.1016/j.nima.2013.03.017}
  {\path{doi:10.1016/j.nima.2013.03.017}}.

\bibitem{Baudot_VCI2013}
J.~Baudot, A.~Besson, G.~Claus, W.~Dulinski, A.~Dorokhov, et~al., {Optimisation
  of CMOS pixel sensors for high performance vertexing and tracking}, 2013.
\newblock \href {http://arxiv.org/abs/1305.0531} {\path{arXiv:1305.0531}},
  \href {http://dx.doi.org/10.1016/j.nima.2013.06.101}
  {\path{doi:10.1016/j.nima.2013.06.101}}.

\bibitem{Feofilov_VCI2013}
G.~Feofilov, paper presented at this conference.

\end{thebibliography}
\end{document}